\documentclass{kluwer}
\usepackage{graphicx} 
\begin{document}
\begin{article}
\begin{opening}
\title{How to use magnetic field information for coronal loop
identification?}
\author{T. \surname{Wiegelmann}\email{wiegelmann@linmpi.mpg.de}}
\author{B. \surname{Inhester}}
\author{A. \surname{Lagg}}
\author{S.K. \surname{Solanki}}
\institute{Max-Planck-Institut f\"ur Sonnensystemforschung
Max-Planck-Strasse 2, 37191 Katlenburg-Lindau, Germany}
\date{DOI: 10.1007/s11207-005-2511-6 \\
Solar Physics, Volume 228, Issue 1-2, pp. 67-78, 2005}

\runningtitle{Loop identification}
\runningauthor{Wiegelmann et al.}

\begin{ao}
Kluwer Prepress Department\\
P.O. Box 990\\
3300 AZ Dordrecht\\
The Netherlands
\end{ao}

\begin{motto}

\end{motto}
\begin{abstract}
The structure of the solar corona is dominated by the magnetic field  because the magnetic
pressure is about four orders of magnitude higher than the plasma pressure. Due to the high
conductivity the emitting coronal plasma (visible e.g. in SOHO/EIT) outlines the magnetic
field lines. The gradient of the emitting plasma structures is significantly lower parallel
to the magnetic field lines than in the perpendicular direction. Consequently information
regarding the coronal magnetic field can be used
for the interpretation of coronal plasma
structures. We extrapolate the coronal magnetic field from photospheric magnetic field
measurements into the corona. The extrapolation method depends on assumptions regarding
coronal currents, e.g. potential fields (current free) or force-free fields (current
parallel to magnetic field).  As a next step we project the reconstructed 3D magnetic field
lines on an EIT-image and compare with the emitting plasma structures. Coronal loops are
identified as closed magnetic field lines with a high emissivity in EIT and a small
gradient of the emissivity along the magnetic field.
\end{abstract}

\keywords{image processing, magnetic fields, EIT, corona, extrapolations}

\abbreviations{\abbrev{KAP}{Kluwer Academic Publishers};
   \abbrev{compuscript}{Electronically submitted article}}

\nomenclature{\nomen{KAP}{Kluwer Academic Publishers};
   \nomen{compuscript}{Electronically submitted article}}

\classification{JEL codes}{D24, L60, 047}
\end{opening}
\section{Introduction}
\label{sec1} The automatic identification of features in solar images provide an
increasingly helpful tool for the analysis of solar physics data. Previous methods
of solar feature identification used, e.g, contrast, intensity and contiguity
criteria in Ca II K images to identify structures like sunspots and plage
\cite{1998ApJ...496..998W,2001SoPh..202...53P}. Here we are mainly interested in
identifying closed loop structures in active regions. Coronal plasma loops are
optical thin and coronal images, e.g. in X-rays or EUV radiation, provide only a 2D
projection of the 3D structures. The line-of-sight integrated character of these
images complicates their interpretation. The corresponding problem that loop
structures visible in images are often a superposition of individual loops occurs
particularly in high resolution TRACE images \cite{1999SoPh..187..261S,
2004ApJ...615..512S}, although occasionally individual loops are clearly visible
over a large fraction of their length \cite{2000ApJ...541.1059A}. To get insights
regarding the structure of the coronal plasma it is important to consider the
distribution of the plasma $\beta$ with height. The plasma $\beta$ varies from
$\beta > 1$ in the photosphere over $\beta \ll 1$ in the mid-corona to $\beta > 1$
in the upper corona (See
\citeauthor{2001SoPh..203...71G}(\citeyear{2001SoPh..203...71G}) for details). For
the interpretation of coronal loops it is helpful that they are located in the low
$\beta$ corona and consequently their structure is dominated by the coronal magnetic
field. In the following we will therefore discuss how information about the magnetic
field can be used for the analysis of coronal images, e.g., from SOHO/EIT,
SOHO/SUMER, or in future from the two STEREO spacecrafts.

The strongest emissivity in  EUV-images originates from closed magnetic
loops in active regions. Due to the high conductivity the coronal plasma
becomes trapped on close magnetic field lines and consequently the
emitting plasma outlines the magnetic field
\cite{1975SoPh...44...83P,1977SoPh...52..397S,1989SSRv...51...11S}.
This feature has previously
been used (in combination with line of sight photospheric magnetograms) to
compute free parameters (e.g., the linear force-free parameter $\alpha$)
in coronal magnetic field models.
\citeauthor{2004ApJ...615..512S}(\citeyear{2004ApJ...615..512S}) used global potential
fields and scaling laws to fill the coronal magnetic field lines with plasma and
computed so called synthetic EIT images from their magnetic field model, which
they compared with real EIT observations. While potential fields give
a rough impression regarding the global structure of the solar corona
(e.g. the location of coronal holes and active regions,
 \citeauthor{2002JGRA.107lSSH13N} \citeyear{2002JGRA.107lSSH13N})
they often fail to describe magnetic fields in active regions accurately.
Here we use different magnetic field models
(potential, force-free fields and coronal magnetic fields
reconstructed directly from measurements)
in combination with coronal plasma images to identify coronal structures,
where we concentrate on closed loops in an active region.

While the space in the corona is basically filled with magnetic field, only some
fraction of all field lines shows a strong emissivity. Our aim is to identify out of
the many (theoretical infinite) number of field lines the ones which are accompanied
by a strong plasma radiation.
\citeauthor{1999ApJ...515..842A} (\citeyear{1999ApJ...515..842A}) picked out
individual loops from two EIT images and applied a geometrical fit (assumption of
circular loops) and dynamic stereoscopy to draw conclusions regarding the 3D
structure of the loops. \citeauthor{2002SoPh..208..233W}
(\citeyear{2002SoPh..208..233W}) used these circular 3D loops
 and a MDI magnetogram to compute the optimal
linear force-free field consistent with these observations.
The method was extended by
\citeauthor{2003SoPh..218...29C} (\citeyear{2003SoPh..218...29C})
to work with 2D coronal images, in which an individual loop  is visible in coronal images along it's
entire length (including the approximate location of both footpoints).
An alternative approach was taken by
\citeauthor{2004A&A...428..629M} (\citeyear{2004A&A...428..629M}),
who identified several loops or (large) parts of
loops in EIT-images (by manual tracing).
The purpose of these papers was not feature identification, but to use
coronal images and line of sight magnetograms to compute the coronal magnetic
field within a linear force-free model. A manual selection of the loops
was necessary.

The aim of this work is to automize these procedures
and allow more general magnetic field models.
We use linear and non-linear
force-free magnetic field models and also give an example with measured
magnetic fields. In the special case of linear force-free fields our method
automatically calculates the optimal linear force-free parameter $\alpha$.

In this sense the method is a further development for computing linear force-free
fields from line-of sight magnetograms and coronal images.
We outline the paper as follows: In section \ref{sec2} we
outline the loop identification method and in section \ref{sec3}
we describe how our method finds the optimal values for magnetic
field models with free parameters. Section \ref{sec4} contains an
example where we tested our algorithm with different magnetic field
models. Finally we draw conclusions and give an outlook for future
work in section \ref{sec5}.
\section{The loop identification method}
\label{sec2}
\begin{figure}
\includegraphics[clip,height=14cm,width=16cm]{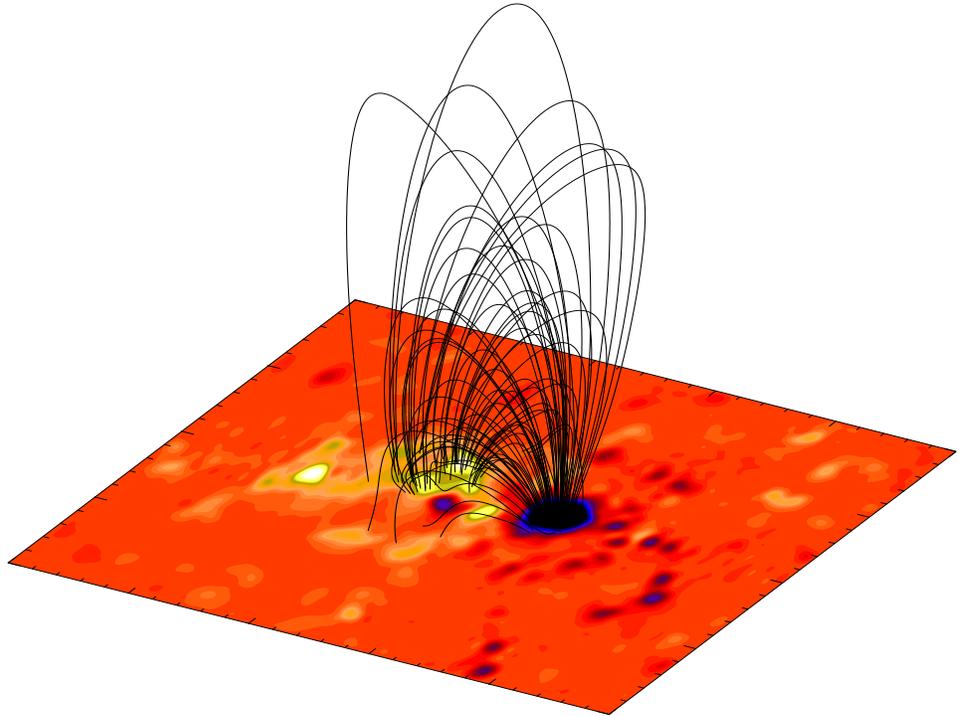}
\caption{3D magnetic field of AR7953. We used a linear force-free model with
$\alpha= -0.11 Mm^{-1}$ to extrapolate photospheric magnetic field measurements into
the corona. We plot only closed magnetic field lines with a magnetic field strength
above $100 G$ at the solar surface. Please note that the vertical scale is stretched
by a factor of two.} \label{fig1}
\end{figure}
\begin{figure}
\mbox{\includegraphics[clip,height=6cm,width=8cm]{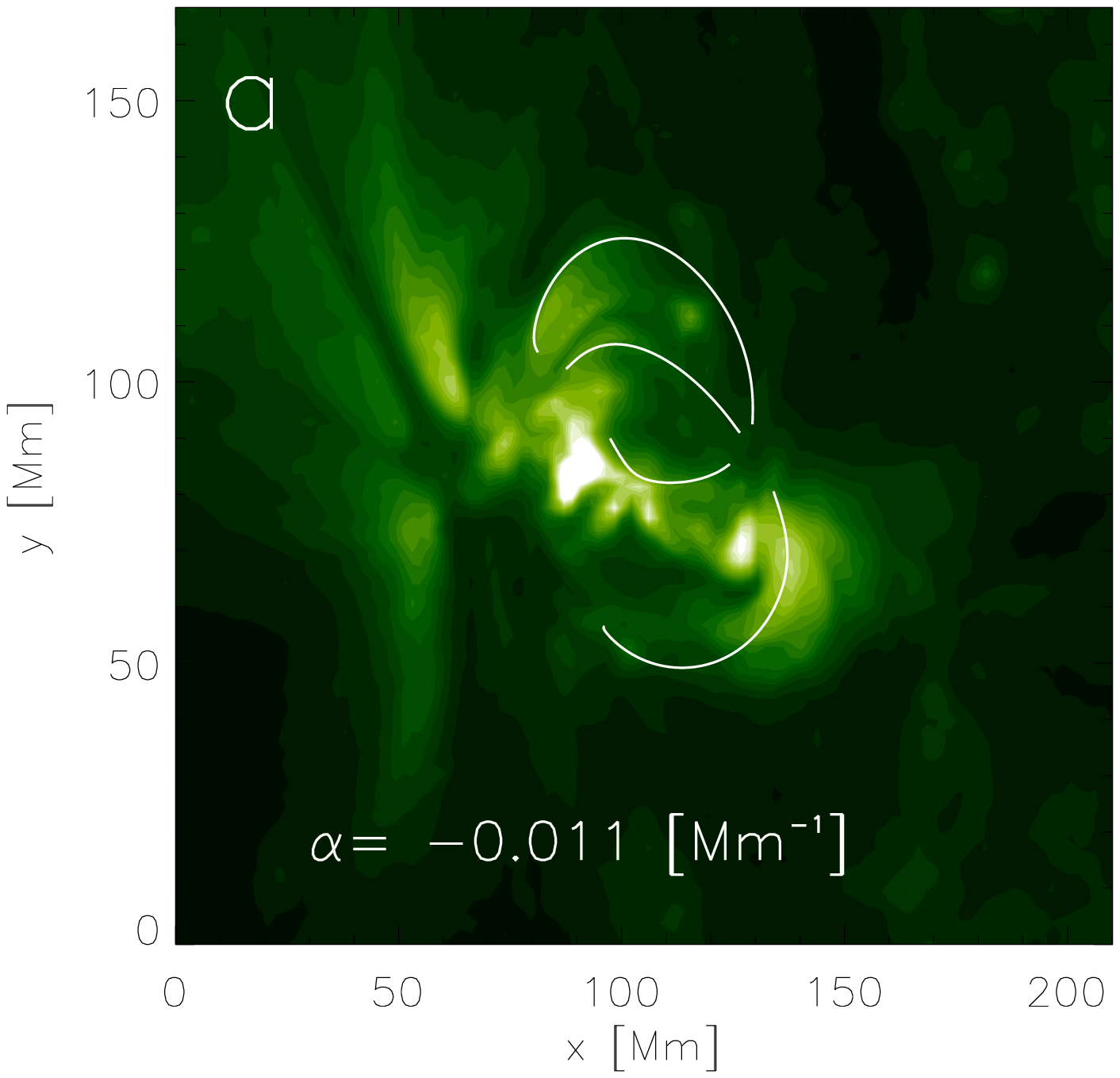}
\includegraphics[clip,height=6cm,width=8cm]{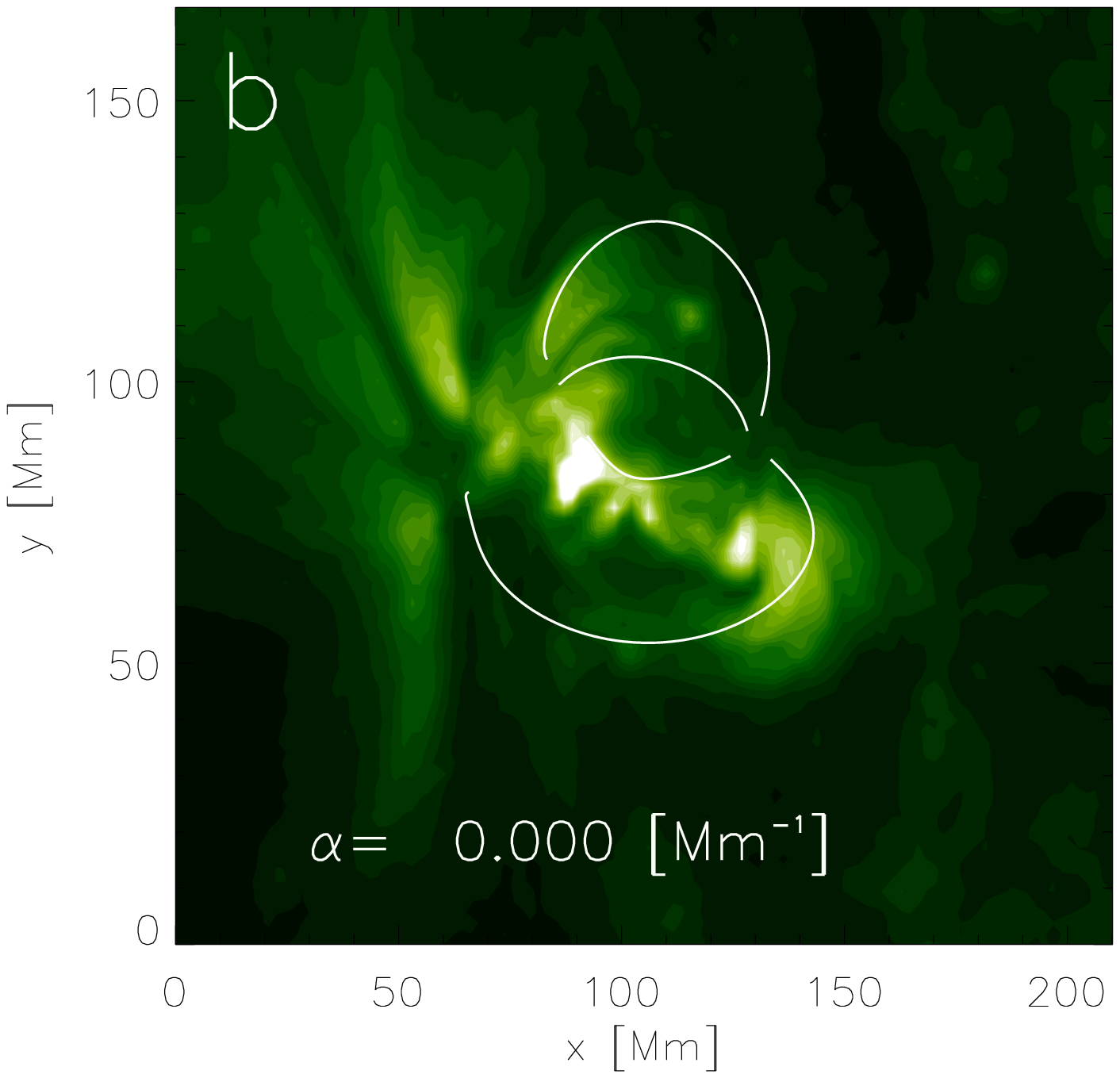}}
\mbox{\includegraphics[clip,height=6cm,width=8cm]{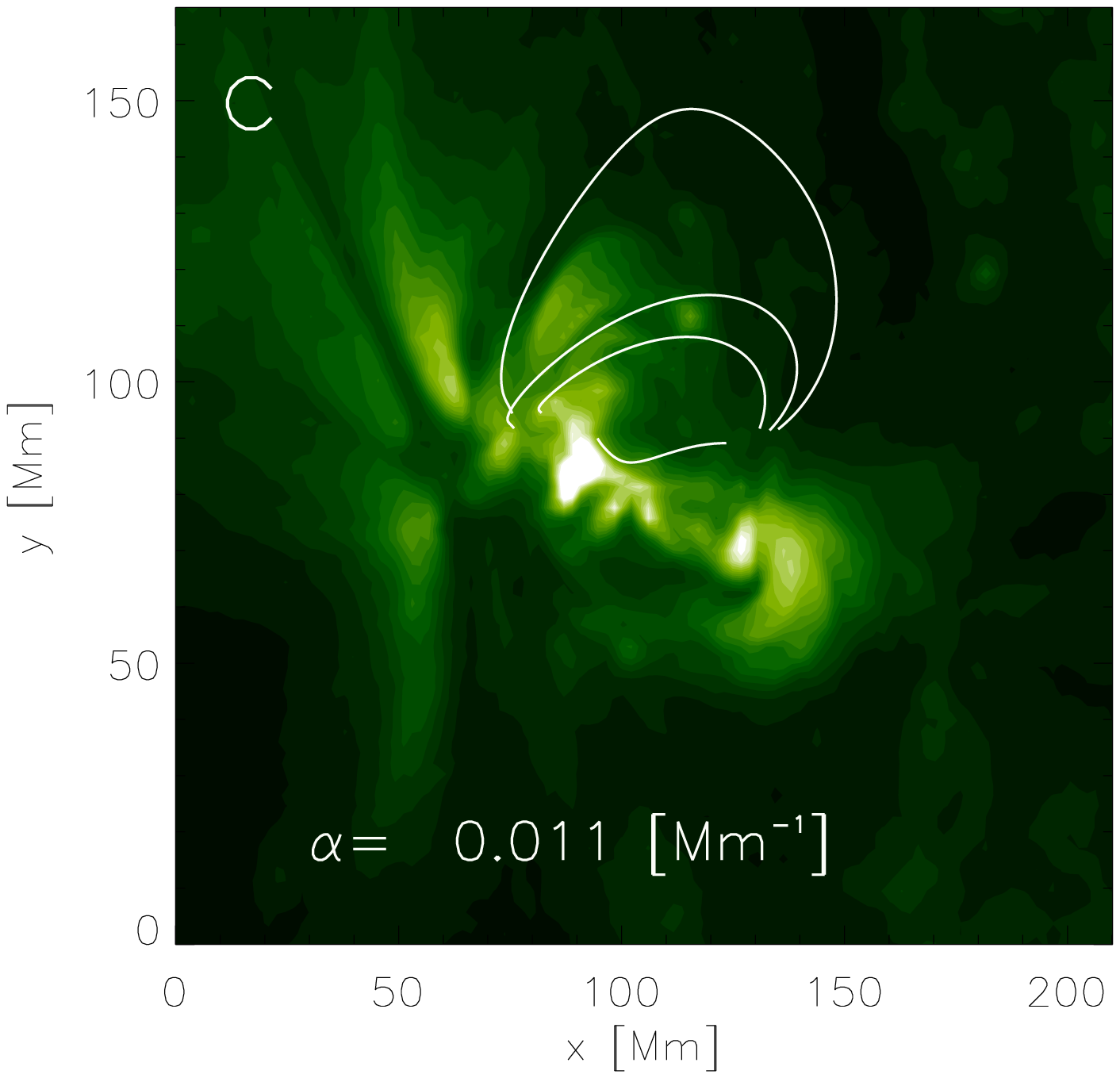}
\includegraphics[clip,height=6cm,width=8cm]{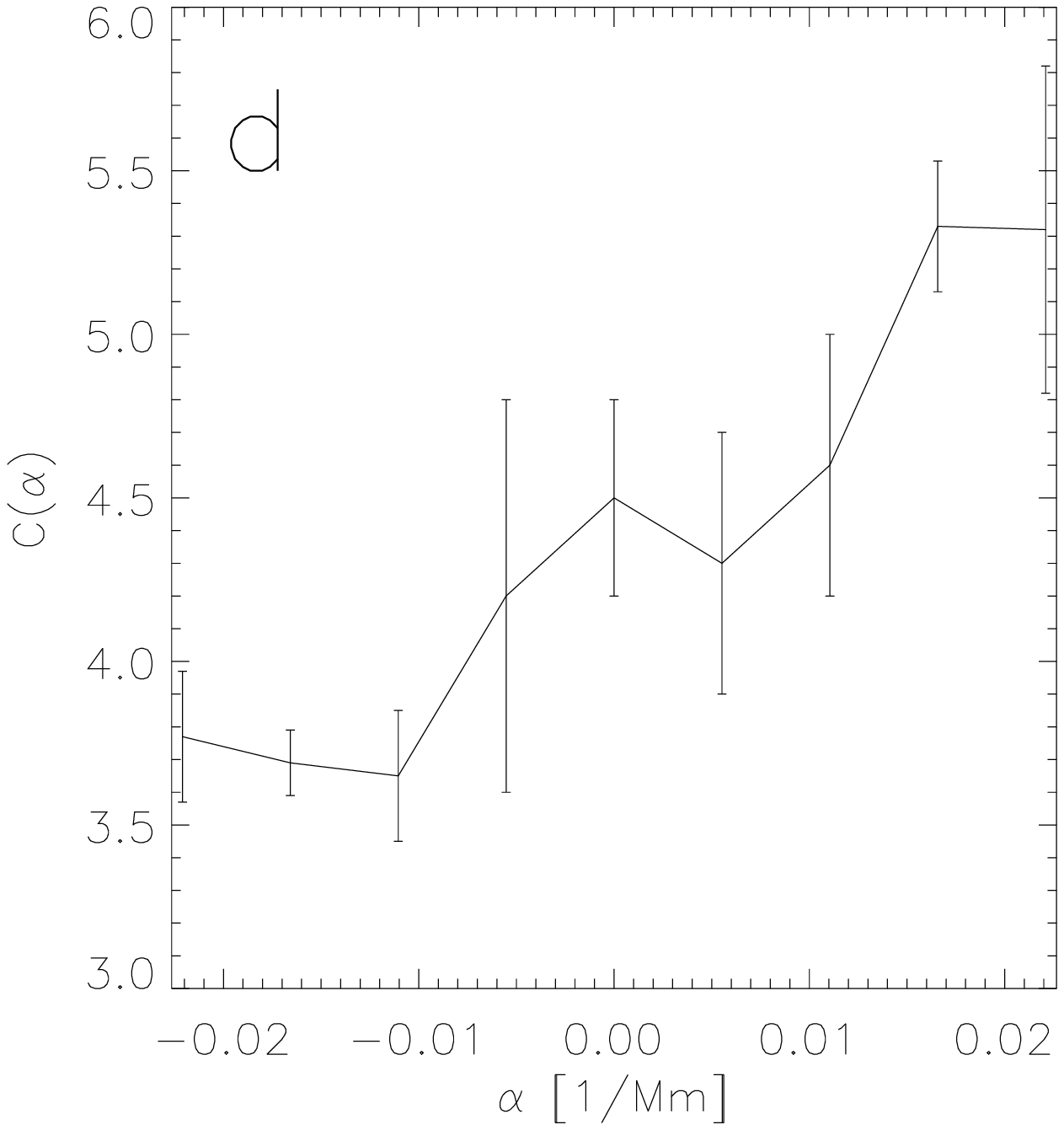}}
\caption{EIT picture of AR 7953 and projections of linear force-free magnetic field
lines. The field lines have been computed with the help of our loop identification
method. It is obvious from the pictures that a linear force-free model with a
negative value of $\alpha=-0.011 Mm^{-1}$ (panel a) fits the EIT-image better than a
potential field (panel b) or a linear force-free field with a positive value of
$\alpha=0.011 Mm^{-1}$ (panel c). Panel d) shows quantitative (value of $C$, see
text) agreement of the magnetic field model and EIT-image. The value of $C$ was
averaged over the best four individual loops. The error-bars indicate the scatter
(standard deviation of $C$ with respect to the four best loops) of the identified
structures with respect to the measure $C$. } \label{fig2}
\end{figure}
Previous loop identification methods used the enhanced brightness of coronal loops
compared with the background.
 \citeauthor{1992PASJ...44L.181K}(\citeyear{1992PASJ...44L.181K}) and
\citeauthor{2000SoPh..193...53K}(\citeyear{2000SoPh..193...53K}) identified loops by
visually selecting the brightest pixels in coronal images and fitting them by least
square with a polynomial function. \citeauthor{1999ApJ...515..842A}
(\citeyear{1999ApJ...515..842A}) subtracted a smoothed image (using a $3 \times 3$
boxcar) from the original to enhance the loop structure. The key assumption for our
method is that the coronal plasma, visible e.g. with EIT, also outlines the coronal
magnetic field. Due to the low plasma pressure in the corona the magnetic field
dominates the structures of the corona and due to the high conductivity the coronal
plasma is trapped on field lines. The method involves the following steps:
\begin{enumerate}
\item Compute the coronal magnetic field.
\item Calculate magnetic field lines.
\item Project all magnetic field lines onto an image of the coronal plasma, e.g. from EIT.
\item Compute emissivity and emissivity gradient of the image along the projected field lines.
\item Check how well projected field lines and plasma features agree
expressed by a quantitative parameter.
\item Sort loops and choose the best ones in the sense of agreement with the plasma features.
\item Plot some representative loops onto the image.
\end{enumerate}
\subsection{Compute the coronal magnetic field.}
Direct observations of the coronal magnetic field are only available for a few
individual cases (see section \ref{sec4}) and usually one has to extrapolate the
coronal magnetic field from photospheric measurements. Full disc line of sight
magnetograms are routinely observed with, e.g., SOHO/MDI and at Kitt Peak.
Assumptions regarding the coronal current flow are essential for the extrapolation
method.
\begin{itemize}
\item Current free potential fields can be computed from LOS-magnetograms alone.
Potential fields are usually not a good model for magnetic fields in active regions.
\item Linear force free fields are also computed from LOS-magnetograms, but contain
a free parameter $\alpha$. We will illustrate later, how our method finds the
optimal value of $\alpha$, i.e. the one fits the observations best. \item Nonlinear
force-free fields require photospheric vector magnetograms as input and are
numerically more expensive to compute. If these data are available, nonlinear
force-free fields are preferred over linear force-free fields. The reason is that
linear force-free fields  are a subclass of non-linear force-free fields and
consequently non-linear fields provide a larger class of solutions.
 Unfortunately, current vector magnetograms usually have a limited field of
view, large noise in the transversal field component and are simply often not
available for active regions of interest. This situation will improve once SOLIS
starts providing regular vector magnetograms.
\end{itemize}
 Let us remark that the result of the extrapolation is as well influenced by the
 lateral and top boundary conditions of the computational box. A popular choice
 for the upper boundary of global potential fields is a source surface
 (all field lines are forced to become radial on the source surface.) The
 extrapolation method of linear force-free fields implicitly specifies the
 upper boundary by assuming that the magnetic field is bounded at infinity
 (see \citeauthor{1978SoPh...58..215S}(\citeyear{1978SoPh...58..215S}) for
 details.) The limited field of view of magnetograms effect the extrapolation as
 well, because distance sources become neglected and one has to specify lateral
 boundary conditions, e.g. symmetry assumptions
 or a boundary layer  towards
 the lateral and top boundaries of the computational box. The effects of the lateral
 boundary conditions on the reconstruction can be diminished by choosing
 a well isolated active region with low magnetic flux close to the side boundaries
 of the computational box
(See \citeauthor{2004SoPh..219...87W}(\citeyear{2004SoPh..219...87W} for details).
Our method usually uses linear force-free fields, but we will present also one
example where we used non-linear force-free fields and coronal magnetic loops
deduced directly from spectropolarimetric observations.

\subsection{Calculate magnetic field lines.}
We use a forth order Runge-Kutta field line tracer to compute a large number of
magnetic field lines. The starting points for the field line integration are chosen
randomly at the solar surface, whereby the number of traced field lines
starting from a given surface area is proportional to the strength of the
magnetogram signal.
The particular concentration on loops with high magnetic flux is motivated by
scaling laws \cite{2004ApJ...615..512S} which suggests that loops with
a strong magnetic field are also bright loops.

Because the coronal plasma is trapped on closed magnetic loops and has a high
emissivity there we are interested here in studying closed magnetic loops.
We therefore  store only field lines which are closed.

Fig. \ref{fig1} shows an example. We computed the 3D magnetic field of AR7953 with
the help of a linear force-free model using the method of
\citeauthor{1978SoPh...58..215S} (\citeauthor{1978SoPh...58..215S}) for a linear
force-free value of $\alpha=-0.011 {\rm Mm}^{-1}$. We plot only closed magnetic
field lines where at least one footpoint has $B_z > 100 G$ in the photosphere. The
value of the magnetic field is implicit also a threshold for the brightness of the
loops and we are interested mainly in identifying bright loops. This is in
particular important for the STEREO mission, where we want to identify pairs of
loops in two images.
\subsection{Project all magnetic field lines onto an image of the coronal plasma, e.g. from EIT.}
The faint, optically thin, structure of the coronal emission does not give
information regarding its height in the corona. Therefore the 3D magnetic field
lines are projected onto an EIT-image on the disk for comparison. It is assumed that
the source of high emission is located in closed magnetic structures.
\subsection{Compute emissivity and emissivity gradient of the image along the projected field lines.}
We compute the emissivity $\rho$ and it's gradient
$\nabla \rho$ along the magnetic field lines. The coronal plasma is assumed
to outline the magnetic field and consequently the emissivity gradient parallel
to the field lines should be small.
\subsection{Check how well projected field lines and plasma features agree.}
\label{sec2.5}
We use the following measure to get quantitative information on how
good a given field line and the plasma emission agree.
\begin{equation}
C=\frac{\int_0\limits^l |\nabla \rho| \; ds}{l \; \left(\int_0\limits^l \rho \; ds \right)^2  }
\label{defc}
\end{equation}
where $l$ is the length of the loop and the integrals sum along the loop. The lower
the value of $C$ the better magnetic loop and plasma agree with each other. One may
certainly play around with the exact parameters in Eq. \ref{defc}, e.g. use simply
$\frac{\int_0\limits^l |\nabla \rho| \; ds}{\int_0\limits^l \rho \; ds}$ or
$\frac{\int_0\limits^l |\nabla \rho| \; ds}{l \; \int_0\limits^l \rho \; ds}$. The
form used in Eq. \ref{defc} gives stronger weight to loops with a high emissivity
and long loops. We compute a value of $C$ for all field lines. \footnote{Let us
remark that the absolute value or normalization of $C$ is unimportant. Different
images with a different emissivity, wavelength etc. will usually have different
ranges for $C$. We multiply the absolute values of $C$ with $1.0e8$ for the examples
presented here to get handy numbers.} The gradient along the center of the coronal
loop is low but also the gradient could be low far from the loops and the
denominator normalization factor may not always correct this if the first potence of
the emissivity is used in the nominator. We find that taking the square of the
emissivity in the nominator ensures that C becomes large far from the loops and the
nominator compensates the low emissivity gradient in the background plasma.  The
computation of $C$ might become influenced by overlapping and crossing loops which
locally enhance the emissivity and the emissivity gradient.

 Since any coronal image samples only gas
within a limited range of temperatures, this method assumes that coronal loops have
a homogeneous temperature along their entire length. This is a relative good
approximation (e.g. \citeauthor{2002A&A...383..661B}
\citeyear{2002A&A...383..661B}).

Another possibility to check the agreement between magnetic field lines
and the coronal emission is to fit a Gaussian to the coronal plasma perpendicular
to the magnetic field lines and to compute the shift
between the center of the Gaussian and magnetic
field lines integrated along the loop \cite{2003SoPh..218...29C}. This method
was especially successful for single isolated loops, where both footpoint areas of the loop
have been identified and only field lines connecting both footpoint areas have been
considered for the Gaussian fit.
Note that this technique does not rely on the assumption of loops with
uniform temperature.
Within this work we are planning to automatically detect
the loops without any prior footpoint identification.
Often one or both footpoints are not visible in the coronal emission. A Gaussian
fit can also give
misleading results when different coronal loops lie close together in the image.
\subsection{Sort loops and choose the best ones in the sense of agreement with the plasma features.}
We sort all field lines with respect to the measure $C$.
In principle every plasma feature consists of an infinite number of magnetic field lines
and after sorting the loops with respect to $C$ it might happen that the best
identified loops (lowest values of $C$) lie close together. For the identification and
illustration of coronal loops it seems to be better that every feature is only represented
by one field line. To do so we compute the difference of all projected field lines
(e.g. the area between two field lines) and choose only loops which are clearly
separated (area between the loops above a certain threshold).
\subsection{Plot some representative loops onto the image.}
Eventually we plot the projections of the best loops onto the coronal images.
Fig. \ref{fig2} a) shows four identified loops. The alignment of the coronal structures with the
magnetic loops is obvious. A more detailed discussion of this figure
is given in section \ref{sec3}.
Let us remark that our method not only provides the locations of the 2D projection
of the loop, but additional information as well. The magnetic field automatically
provides the 3D structure of the identified coronal features, the magnetic
field strength and current density along the loop. Together with other information
from coronal images, e.g. information regarding the plasma flow with SUMER, we
can draw further conclusions regarding physical properties. For example
\citeauthor{2004A&A...428..629M} (\citeyear{2004A&A...428..629M})
studied plasma
flows along magnetic field lines for this active region with the help of SUMER data.
\section{Using magnetic field models with free parameters}
\label{sec3}
 Coronal images give information regarding the direction of the coronal magnetic
 field. A magnetic field model provides the 3D-magnetic field in the
 entire space and projections of corresponding magnetic field lines onto
 the disk can be compared with the coronal image. For
 non-linear force-free field extrapolations from vector magnetograms
 or models based on measurements of coronal magnetic fields the projected
 model field lines are independent from the coronal image. Magnetic field model
 and coronal image provide independent from each other information regarding
 the direction of the magnetic field.
 Unfortunately models based on non-linear force-free extrapolations or direct
 measurements are
currently only available for a few individual cases. Usually the only available
magnetic field data are line-of-sight magnetograms from e.g. MDI or Kitt Peak. Under
the assumption of current-free potential fields the coronal magnetic field can be
derived from LOS-magnetograms. Unfortunately current-free models are usually not a
good description for magnetic fields in active regions. We can, however, consider
electric currents with the help of linear force-free models where the current free
parameter $\alpha$ is unknown a priori and cannot be derived from the
LOS-magnetogram. \citeauthor{2002SoPh..208..233W} (\citeyear{2002SoPh..208..233W})
used 3D stereoscopic reconstructed loops from \citeauthor{1999ApJ...515..842A}
(\citeyear{1999ApJ...515..842A}) to fix $\alpha$ and
\citeauthor{2003SoPh..218...29C} (\citeyear{2003SoPh..218...29C}) fitted $\alpha$
with the help of a plasma loop identified earlier.

We generalize the method and use the  EIT-image (without any pre-identification of
structures) together with the fitting routine described in section \ref{sec2} to
compute the optimal value of $\alpha$. To do so, we compute magnetic field lines for
several values of $\alpha$ and apply the feature identification method (see Sect.
\ref{sec2}) to each of the models and compute the values of $C$ for several loops in
each model. As an example we tried to identify coronal loops with the help of a
potential field model $\alpha=0$ in Fig. \ref{fig2} b) and with $\alpha=0.01 {\rm
Mm}^{-1}$ in Fig. \ref{fig2} c). It is obvious from the images that the alignment of
plasma structures and magnetic field lines is worse than for the optimal value of
$\alpha=-0.01 {\rm Mm}^{-1}$. The value of $\alpha$ corresponding to the lowest
value of $C$ depicts the optimal value of $\alpha$. As a consequence the method
automatically computes the optimal coronal magnetic field (within the linear
force-free approximation) from LOS-magnetograms and a plasma image (within the
limitations listed in Sect. \ref{sec2.5}). In Fig. \ref{fig2} d) we show
quantitatively, with help of the measure define in Eq. \ref{defc} how well magnetic
field model and emission structures agree with respect to $\alpha$. The minimum of
the function depicts the optimal value of the linear force-free parameter,
$\alpha=-0.01 {\rm Mm}^{-1}$.
 Let us remark that the main reason for using linear-force free fields are their
 mathematical simplicity and because non-linear force-free models require
 vector magnetograms which are often not available
or have a limit FOV and high noise/ambiguity in the transversal field. If these data
are available a non-linear force-free model is preferred over a linear one. We find,
however, that linear force-free fields often give a reasonable approximation of the
coronal magnetic field in active regions (much better than potential fields. See
section \ref{sec4} and \cite{twinpress}).
 The linear force-free model  has only one free
parameter, but in principle one can also use a magnetic field models with several
free parameters, e.g. a wider class of magnetohydrostatic equilibria
\cite{2000A&A...356..735P}, together with a multi-dimensional search for the optimal
(lowest value of $C$) parameter set.
\section{Using different magnetic field models}
\label{sec4}
\begin{figure}
\mbox{\includegraphics[clip,height=6cm,width=8cm]{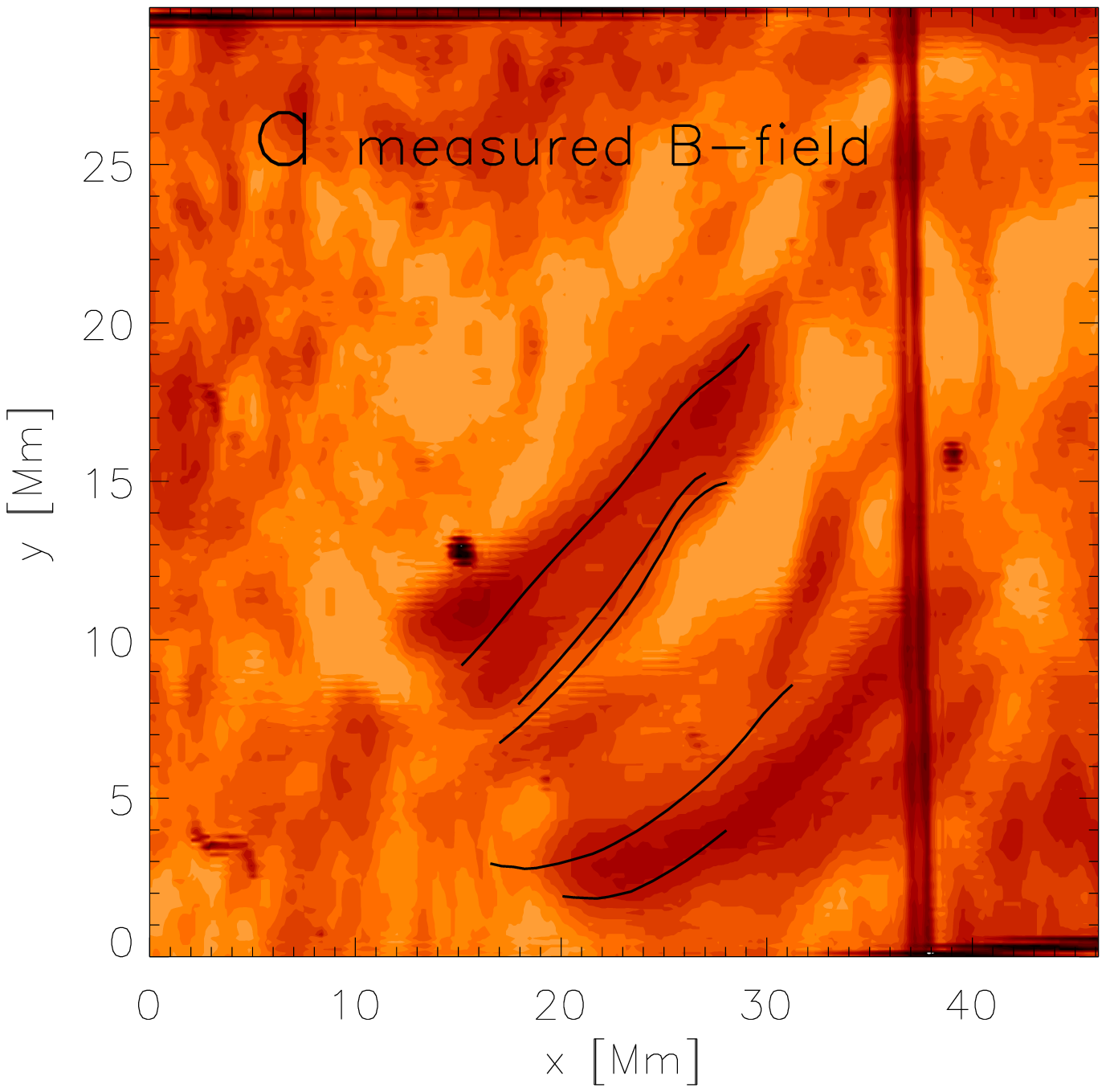}
\includegraphics[clip,height=6cm,width=8cm]{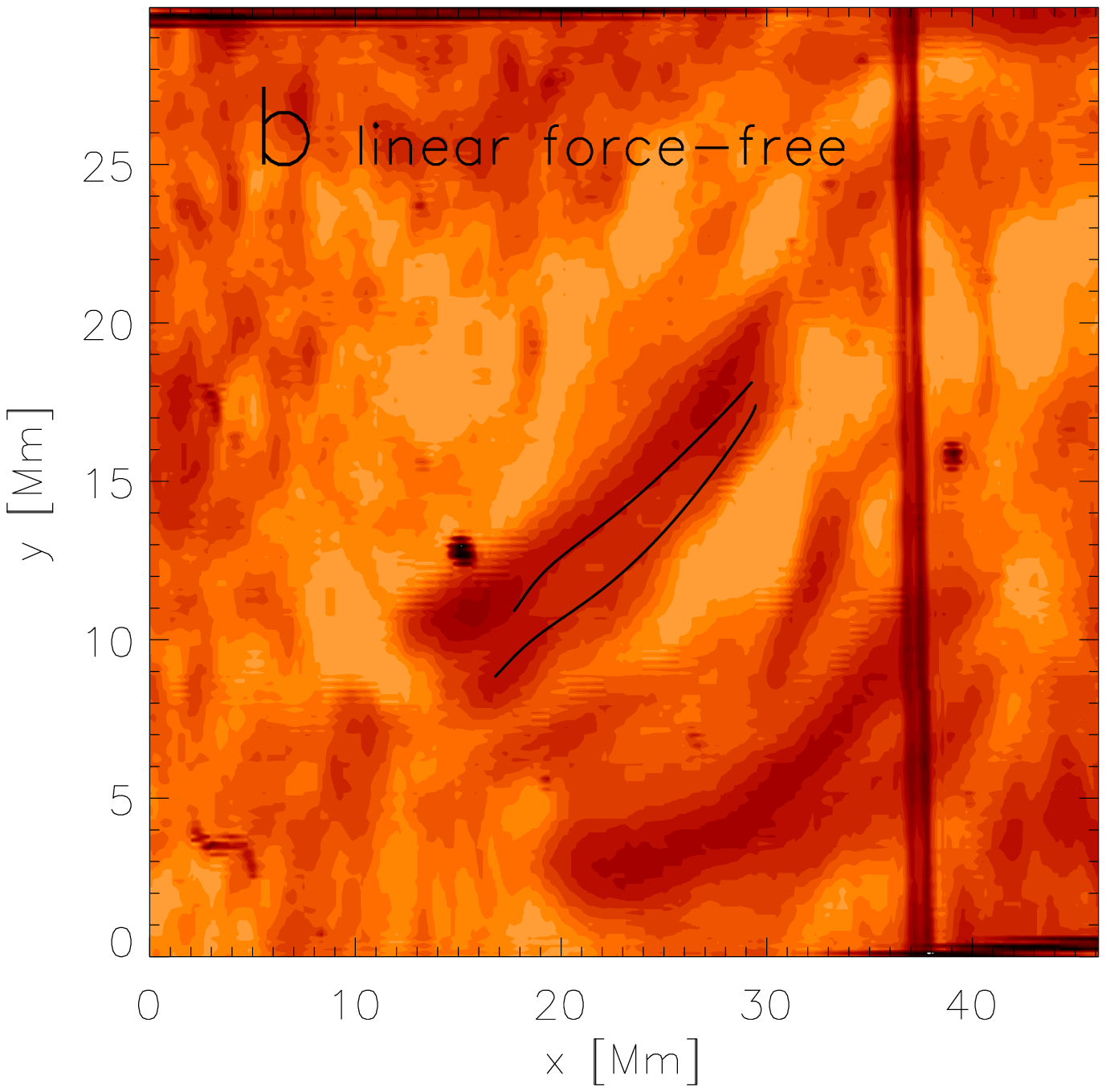}}
\mbox{\includegraphics[clip,height=6cm,width=8cm]{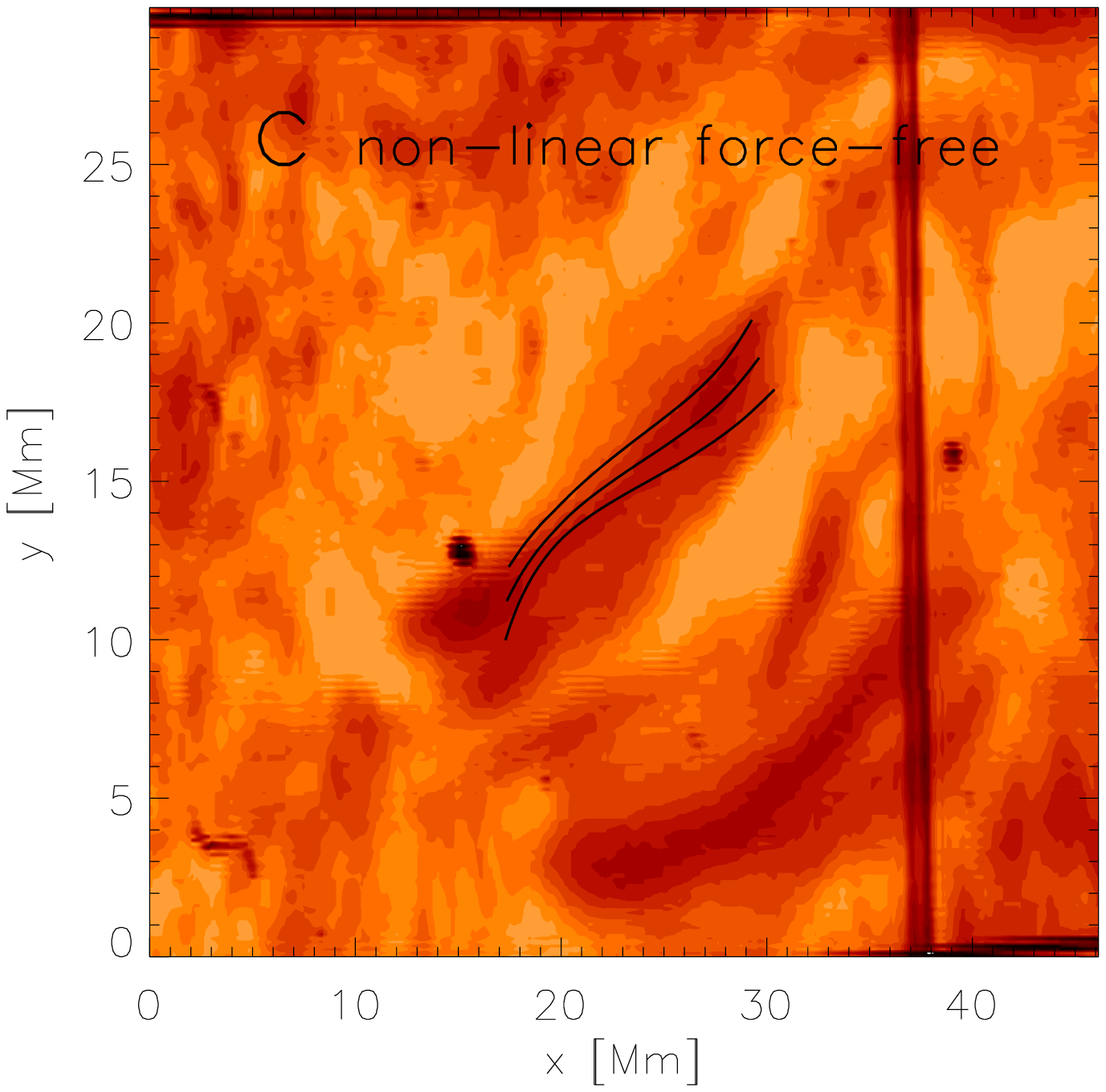}
\includegraphics[clip,height=6cm,width=8cm]{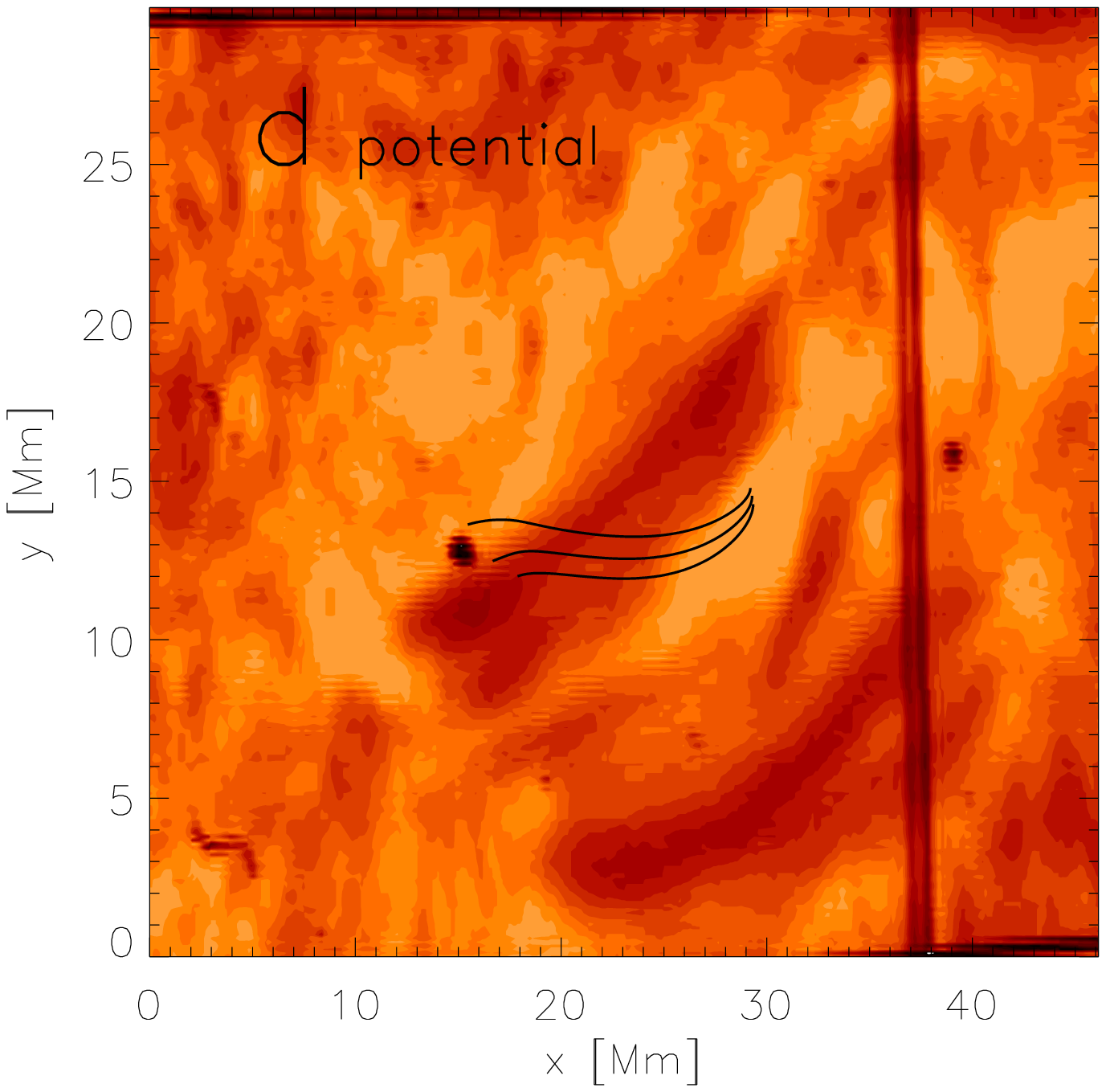}}
\caption{Identification of loops in H Alpha image with
different magnetic field models
(a) measured field, (b) linear force-free field with
$\alpha=0.085 {\rm Mm}^{-1}$, (c) nonlinear force-free
field, (d) potential field. We find that all used magnetic
field models, except the potential field, identify the
loop in the center of the image. The values of $C$ for this loop
are a) $1.57$, b) $1.34$, c) $1.34$ and for potential fields almost
one order of magnitude worse d) $10.78$.
The extrapolated fields (b-c) cannot detect the loop structure in
the lower part of the image because of the limited field of view
of the vector magnetogram. We applied our identification
procedure to the inverse image, because H Alpha is an absorption
line.}
\label{fig3}
\end{figure}
Until now we used linear force-free models for the loop identification
and in many cases this will remain the first choice because
this model  requires only line-of-sight magnetograms, which are routinely
observed and  available online
(MDI: http://soi.stanford.edu/, Kitt Peak: http://www.noao.edu/kpno/).
In the near future, when vector magnetograms will
be observed routinely (e.g. from SOLIS or Solar-B)
 non-linear force-free models can be employed.

Here we provide an example of our fitting routine with different magnetic field
models. We use a vector magnetogram of the developing active region NOAA 9451
observed with the German Vacuum Tower Telescope (VTT) and the Tenerife Infrared
Polarimeter. The pictures have been taken close to the disk center
$\mu=cos(\theta)=0.8$ and the images have not been corrected. We compute a potential
field and a linear force-free field from the line-of sight component of the
magnetogram using the method of \citeauthor{1978SoPh...58..215S}
(\citeyear{1978SoPh...58..215S}) (See Figs. \ref{fig3} b) and d), respectively).
With the help of an optimization principle
\cite{2000ApJ...540.1150W,2004SoPh..219...87W} we also compute a non-linear
force-free field from the full vector magnetogram (See Fig. \ref{fig3} panel c). For
this active region the chromospheric magnetic field vector has been measured
\cite{2003Natur.425..692S,2004A&A...414.1109L} (See Fig. \ref{fig3} a). The direct
field measurements as well as the linear and non-linear extrapolation methods (Fig.
\ref{fig3} panels a-c, respectively) are able to identify a loop structure in the
center of a corresponding H Alpha image with similar accuracy $C=1.4 ^+_- 0.1$. The
potential field is not able to identify any structure in the image and the measure
$C=10.78$ is almost an order of magnitude higher than for the other magnetic field
models. This result is not surprising, because the direct magnetic field
measurements have been compared with extrapolated magnetic fields and the comparison
revealed that a potential field model was not appropriate for this developing active
region \cite{twinpress}. The comparison with different magnetic field models has
revealed that field line parallel electric currents are important to describe the
structure and thus the configuration contains magnetic shear.
\section{Conclusions and outlook}
\label{sec5}
Within this work we have developed a method for the automatic
identification of coronal loops. The method uses a suitable
coronal magnetic field model, e.g. from a force-free extrapolation,
and compares the projection of 3D magnetic field lines with structures
on images, e.g. from SOHO/EIT. As a result coronal structures are
represented by 1D curves. A further use of the method is planned in
particular for the analysis of data from the two STEREO spacecrafts.
The reconstruction of curves is attractive for the STEREO mission for two
reasons. Firstly structures on the solar surface are dominated by thin flux
tubes which often can well be approximated by idealized curves.
Secondly, the stereoscopic reconstruction problem for 1D objects is well
posed. Formally, a projected curve in an image can be projected back along
the view direction giving rise to a 3D solution surface on which the
3D curve must lie. From two images, we obtain two such surfaces and
their intersection must be the solution curve.
This is different for a point-like object. Its projection yields two
coordinate parameters in two images while only 3 spatial coordinates
are required. Formally, the triangulation of point-like objects is therefore
over-determined. Conversely, if we want to solve for optically thin,
curved surfaces we have an underdetermined problem because the bright limb
visible in the two images might project back to different parts of the
surface. It is therefore useful to identify 1D objects (loops)
out of 2D images for a stereoscopic
reconstruction by e.g. tie-point-like methods \cite{2004AGUFMSH21B0422H}.
Let us remark that the structure identification method by magnetic field
lines provides already a 3D structure of the loop which might be further
refined by stereoscopic methods.
In particular, the advent of the STEREO mission has produced an increased interest of
the solar physics community in stereoscopic
(\citeauthor{1999ApJ...515..842A},\citeyear{1999ApJ...515..842A},\citeyear{2000ApJ...531.1129A}
and tomographic techniques
\cite{1999JGR...104.9727Z,2000ApJ...530.1026F,2003SoPh..214..287W}
used in other branches of physics, image and information sciences.
However, unlike many of the ``experiments'' conducted in the latter
disciplines, the STEREO mission does not provide sufficient control over
many of the parameters which determine the conditions for an optimal
3D reconstruction. Tomographic methods will in particular be useful in
situations where individual loops cannot be identified due to the
superposition of many loops due to the line-of-sight integrated character
of 2D images.
\begin{acknowledgements}
We acknowledge the use of data from SOHO/EIT, NSO/Kitt Peak and the German VTT. The
work of T. Wiegelmann was supported by  DLR-grant 50 OC 0007 and by a JSPS visitor
grant at the National Astronomical Observatory in Tokyo.
\end{acknowledgements}

\end{article}
\end{document}